\documentclass[a4paper,11pt]{revtex4}
\usepackage{graphicx} 
\usepackage{amsmath} 
\usepackage{amssymb} 
\usepackage{bm} 
\usepackage{dcolumn}
\usepackage{color}
\usepackage{mathrsfs}
\usepackage{amsfonts}
\usepackage{varioref}
\usepackage{mathrsfs}
\usepackage{graphicx}
\usepackage{latexsym}
\usepackage{amsmath}
\usepackage{amssymb}
\usepackage{textcomp}
\usepackage{amsbsy}
\usepackage{graphics}
\usepackage{epstopdf}
\usepackage{color}
\usepackage[caption=false]{subfig}

\RequirePackage[colorlinks,citecolor=blue,urlcolor=magenta,linkcolor=blue]{hyperref}
\input epsf

\allowdisplaybreaks[4]
\begin{document}

\tolerance=5000

\title{Origin of bulk viscosity in cosmology and its thermodynamic implications}

\author{Tanmoy~Paul$^{1}$\,\thanks{tanmoy.paul@visva-bharati.ac.in}} \affiliation{
$^{1)}$ Department of Physics, Visva-Bharati University, Santiniketan 731235, India}


\tolerance=5000

\begin{abstract}
The purpose of the present work is two folded: (1) we propose a new mechanism for the origin of bulk viscosity in cosmological context, and then, (2) we address the thermodynamic implications of viscous cosmology based on the thermodynamics of apparent horizon. In particular, we show that the velocity gradient of the comoving expansion of the universe (along the distance measured from a comoving observer) in turn leads to a viscous like pressure of the fluid inside the apparent horizon. This points the importance of the bulk viscosity of fluid during the cosmological evolution of the universe, as the comoving expansion itself generates the viscosity. Therefore the thermodynamic interpretations of viscous cosmology becomes important from its own right. In this regard, it turns out that the cosmological evolution of the universe with a bulk viscosity $naturally$ satisfies the first and the second laws of thermodynamics of the apparent horizon, without imposing any exotic condition and for a general form of the coefficient of viscosity. This in turn affirms the thermodynamic correspondence of viscous cosmology.
\end{abstract}


\maketitle

\section{Introduction}
The discovery of black hole thermodynamics associated with event horizon of the black hole puts two apparently different arenas of Physics, namely gravity and thermodynamics, on same footing \cite{PhysRevD.7.2333,Hawking:1975vcx,Bardeen:1973gs,Wald_2001}. In this regard, the entropy of black hole is generally considered to be the Bekenstein-Hawking type entropy (or have some other form), and the temperature of the black hole is given by the surface gravity of the same. Then the thermodynamic law of the black hole results to the underlying gravitational field equations. Consequently, several works have related the black hole thermodynamics with the information theory via the Landauer principle, which in turn affirms the connection between black hole gravity and thermodynamics \cite{Raju:2020smc,Herrera:2020dyh,Song_2007,KIM_2010,Bagchi:2024try}.

In the context of cosmology, the Friedmann-Lema\^{i}tre-Robertson-Walker (FLRW) spacetime acquires an apparent horizon, and similar to black hole thermodynamics, the cosmological field equations can be derived from the thermodynamic laws of the apparent horizon \cite{Jacobson_1995,Cai:2005ra,Cai:2006rs,Paranjape_2006,Nojiri:2024zdu,Nojiri:2025gkq}. For a spatially flat FLRW metric given by,
\begin{eqnarray}
 ds^2 = -dt^2 + a^2(t)\left(dr^2 + r^2d\Omega^2\right) \ ,
 \label{metric}
\end{eqnarray}
(with $t$ is the cosmic time, $a(t)$ represents the scale factor of the universe and $d\Omega^2$ is the line element for a unit 2-sphere surface), the apparent horizon has the radius at,
\begin{eqnarray}
 R_\mathrm{h}(t) = \frac{1}{H(t)} \ .
 \label{app-hor}
\end{eqnarray}
Here $H(t) = \frac{\dot{a}}{a}$ is the Hubble parameter of the universe. It may be noted that, unlike to the black hole thermodynamics, the apparent horizon in cosmological context is dynamical in nature; in particular, $R_\mathrm{h}$ increases with the cosmic time provided the fluid inside the horizon satisfies the null energy condition. Another important quantity is the surface gravity on the apparent horizon: $\kappa = \frac{1}{2\sqrt{-h}}\partial_{a}\left(\sqrt{-h}h^{ab}\partial_{b}R\right)\big|_{R_\mathrm{h}}$ (with $h_{ab} = \mathrm{diag.}(-1,a^2)$ defines the induced metric along constant $\theta$ and constant $\varphi$, i.e. $h_{ab}$ is the induced metric along the normal of the apparent horizon) which, due to the spatially flat FLRW metric, is given by
\begin{eqnarray}
 \kappa = -\frac{1}{R_\mathrm{h}}\left(1 - \frac{\dot{R}_\mathrm{h}}{2}\right) \, ,
 \label{surface gravity}
\end{eqnarray}
which is identified with the horizon temperature, namely,
\begin{eqnarray}
 T_\mathrm{h} = \frac{|\kappa|}{2\pi} = \frac{1}{2\pi R_\mathrm{h}}\left(1 - \frac{\dot{R}_\mathrm{h}}{2}\right) \, .
 \label{temperature}
\end{eqnarray}
Consequently, the growing interests on the interconnection between cosmology and thermodynamics leads to several proposals regarding the form of the entropy of the apparent horizon, such as, the Bekenstein-Hawking like entropy \cite{Jacobson_1995}, the Tsallis entropy \cite{Tsallis:1987eu}, the R\'{e}nyi entropy \cite{Renyi}, he Barrow entropy \cite{Barrow_2020}, the Kaniadakis entropy \cite{Kaniadakis_2005}, the Sharma-Mittal entropy \cite{Sayahian_Jahromi_2018}, the Loop Quantum gravity entropy \cite{Majhi:2017zao}, or more generally, the few parameter dependent generalized entropy \cite{Nojiri:2022aof,Nojiri_2022,Odintsov:2022qnn}. With such different forms of horizon entropies, the entropic cosmology turns out to have a rich implications on describing inflation, dark energy, reheating and bouncing as well \cite{Bousso_2002, fischler1998, Saridakis:2020zol, Sinha_2020, Adhikary:2021,Nojiri:2022aof,Nojiri_2022,Odintsov:2022qnn,Odintsov_2023,Cruz:2023xjp,Luciano:2024bco,Odintsov:2024sbo,Jizba:2024klq,Volovik:2024eni}. However the question that whether the cosmic evolution of the universe is really connected with some thermodynamic laws of the horizon --- still hinges the modern cosmology. One of the ways to understand this issue is to investigate whether the first and the second laws of thermodynamics of the apparent horizon are naturally satisfied in cosmological context. Based on the thermodynamics of apparent horizon, the authors of \cite{Odintsov:2024ipb} recently demonstrated that the cosmic evolution of the universe, governed by the usual Friedmann equations, naturally validates the thermodynamic laws without imposing any exotic condition.

The introduction of viscosity in the context of entropic cosmology to investigate the reversible (or irreversible) expansion of the universe is important from its own right. In the late accelerating universe, the effect of bulk viscous fluid was studied in \cite{PhysRevD.79.103521,BARROW1986335,PhysRevD.101.044010,doi:10.1142/S0218271817300245,PhysRevD.73.043512,PhysRevD.72.023003,Fabris_2006,PhysRevD.82.063507,Avelino_2009,Avelino_2010}. Actually the dark energy fluid requires a negative pressure to achieve an accelerated expansion of the universe, and the bulk viscosity can do this job by providing an effective pressure to the fluid. In general, the analysis of viscous cosmology considers the bulk viscosity of the fluid without affecting the underlying symmetry, i.e. the homogeneity and isotropy, of the background spacetime. This is due to the reason that the effective pressure of the fluid coming from the viscosity depends only on the cosmic time, and thus keeps the homogeneous and isotropy of the spacetime. In this regard, the bulk viscosity has been also applied to unify the dark energy and the dark matter with same fluid, as for instance in the case of the Chaplygin gas \cite{PhysRevD.98.043515,CHIMENTO2005146} or a logotropic fluid \cite{FERREIRA2017213}. In addition, the viscosity terms may play an essential role during the early time inflationary stage \cite{doi:10.1142/S0218271817300245}. Despite such numerous successful implications of viscous cosmology, a serious drawback faced by the bulk viscous fluid is to find a valid mechanism for the origin of bulk viscosity in the expanding universe. According to some theoretical point of view, the bulk viscosity pressure arises when the cosmological fluid expands or contract too fast \cite{PhysRevD.75.043521,PhysRevD.61.023510}, and ceases when the fluid reaches the thermal equilibrium again. However a proper mechanism for the origin of the bulk viscosity still remains as one of the mysteries in cosmology and the search is still on. Moreover the effect of viscosity in entropic cosmology is worthwhile to study, in particular, it is important to investigate whether the cosmological evolution of the universe with bulk viscosity satisfy the first and the second law of thermodynamics of the apparent horizon. This is worthwhile to study as it can establish the thermodynamic correspondence of viscous cosmology.

Based on the above arguments, we try to address the following questions in the present work:
\begin{itemize}
 \item What is the mechanism for the origin of bulk viscosity in cosmology ?

 \item Does the cosmic expansion of the universe with bulk viscosity naturally satisfy the first and the second laws of thermodynamics of the apparent horizon ?
\end{itemize}

It turns out that the comoving expansion of the universe itself is the main agent to generate the bulk viscosity in cosmological context, which immediately points the importance of the bulk viscosity during the cosmological evolution of the universe.

\section{Origin of bulk viscosity in cosmology from Hubble expansion}\label{sec-generation}

\begin{figure}[!h]
\begin{center}
\centering
\includegraphics[width=2.0in,height=2.0in]{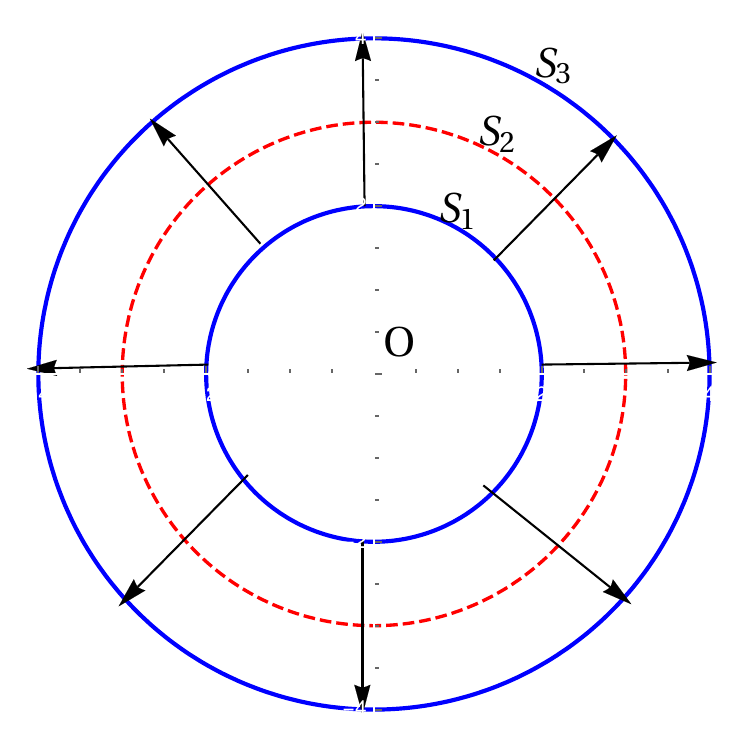}
\caption{'O' is a comoving observer; $S_1$, $S_2$ and $S_3$ are three arbitrary layers in space, having radius as $OS_2 = d$, $OS_1 = d-\lambda_\mathrm{m}$ and $OS_3 = d+\lambda_\mathrm{m}$. Here $\lambda_\mathrm{m}$ represents the mean free path of the fluid inside the horizon.}
 \label{plot-2}
\end{center}
\end{figure}

In this section, we propose a mechanism for the generation of viscosity in cosmology by using the principles of kinetic theory of gas. In particular we show that the velocity gradient coming from the comoving expansion of the universe, i.e. $v = Hd$ (where $H$ is the Hubble parameter of the universe and $d$ is the distance from a comoving observer of a point in space under consideration), is the main reason for generating viscosity in cosmological context. The expression $v = Hd$ clearly indicates that the comoving speed of the universe exhibits a $linear$ gradient along the distance ($d$). Moreover the matter field inside the horizon is considered to be a fluid and obeys the principles of kinetic theory of gas, which has a temperature $T_\mathrm{m}$ and a mean free path $\lambda_\mathrm{m}$ (where the suffix 'm' is for matter fields). Due to such non-zero temperature, the fluid also possesses a random thermal motion, on top of $v = Hd$ coming from the comoving expansion of the universe. Therefore as a whole, the fluid shows two types of motion:
\begin{itemize}
 \item the comoving motion given by $v = Hd$ at a distance $d$ from a comoving observer, which we may identify as mass motion of the fluid,

 \item the random thermal motion due to its temperature $T_\mathrm{m}$.
\end{itemize}
In order to have a clear demonstration, let us consider the Fig.~[\ref{plot-2}] where 'O' is a comoving observer. $S_{1}$, $S_2$ and $S_3$ represent three arbitrary layers in space, $S_2$ lying between $S_1$ and $S_3$ and separated by a distance $\lambda_\mathrm{m}$. If the distance of $S_2$ from the observer is given by $d$, i.e. $OS_2 = d$, then $OS_1 = d-\lambda_\mathrm{m}$ and $OS_3 = d+\lambda_\mathrm{m}$. Due to the thermal motion, the fluid molecules cross the layer $S_2$ from below as well as from above. As a result, along due to the gradient of the mass motion, there exists a net flow of momentum from the layer. We now calculate the rate of such momentum flux, which eventually leads to the viscous pressure in this content.

Let $n$ be the number of molecules that pass through unit area of the $S_2$ from the below layer ($S_1$), per unit time. Therefore, due to the random motion, the same $n$ number of molecules cross the unit area of $S_2$ from the above layer ($S_3$), at unit time. If $\rho$ is the energy density of the fluid, $u$ is the energy/molecule and $\left< c \right>$ is the average thermal speed of the fluid molecule, then
\begin{align}
&\, n= \frac{\rho}{u} \left< c \right> = \frac{\rho}{\frac{f}{2} k_\mathrm{B}T_\mathrm{m}} \left< c \right> \, ,
\label{2}
\end{align}
where $f$ is the dynamical degree of freedom of the fluid molecule and $k_\mathrm{B}$ represents the Boltzmann constant. Moreover, being $\lambda_\mathrm{m}$ is the mean free path of the matter molecule, the $n$ number of molecules have their last collision at $\lambda_\mathrm{m}$ distance away from the layer. Therefore the molecules crossing $S_2$ from $S_1$, have total momentum
\begin{align}
&\, P_\mathrm{below} = n m~v_\mathrm{(d - \lambda_\mathrm{m})} = n m (d - \lambda_\mathrm{m}) H \, ,
\label{3}
\end{align}
where $m$ is the mass of the fluid molecule and we use $v_\mathrm{(d - \lambda_\mathrm{m})} = (d - \lambda_\mathrm{m}) H$, the comoving expansion speed at the distance of $d-\lambda_\mathrm{m}$. Similarly, the molecules passing $S_2$ from $S_3$, carry the following momentum:
\begin{align}
P_\mathrm{above} = n m (d + \lambda_\mathrm{m}) H
\label{4}
\end{align}
Thus $P_\mathrm{below}$ \& $P_\mathrm{above}$ denote the amounts of momentum transformed per unit time \& unit area to the layer $S_2$, in the upward \& in the downward direction, respectively. Therefore the rate of momentum flux from unit area of the layer ($S_2$) from below to above is given by:
\begin{eqnarray}
\frac{dP_\mathrm{flux}}{dtdA} = P_\mathrm{below} - P_\mathrm{above}&=&- 2 nm \lambda_\mathrm{m} H \nonumber \\
 &=&- \frac{\rho \left< c \right>}{f k_\mathrm{B}T_\mathrm{m}} m\lambda_\mathrm{m} H \, ,
 \label{5}
\end{eqnarray}
where we use Eq.~(\ref{2}) to get the second line in the above equation. This can be identified as a viscous pressure on the layer $S_2$, in particular, the viscous pressure on $S_2$ is given by,
\begin{eqnarray}
 p_\mathrm{vis}&=&- \frac{\rho \left< c \right>}{f k_\mathrm{B}T_\mathrm{m}} m\lambda_\mathrm{m} H \nonumber\\
 &=&- 3\zeta \times \frac{dv}{d(d)} \, ,
 \label{N-1}
\end{eqnarray}
where $\zeta =  \frac{\rho \left< c \right>}{3f k_\mathrm{B}T_\mathrm{m}} m\lambda_\mathrm{m}$ and $\frac{dv}{d(d)} = H$ giving the gradient of the mass motion (note that the momentum terms are symbolized by 'P', and the pressure terms are represented by 'p'). Here $\zeta$ acts as the coefficient of viscosity. Therefore we may argue that the comoving expansion of the universe generates a viscous like pressure of the fluid inside the apparent horizon in cosmological context. In this regard, the following points need to be mentioned: (1) due to the negative sign in Eq.~(\ref{N-1}), $p_\mathrm{vis}$ on a layer, exerted by another layer, is along the opposite direction of the relative motion of the layers; and (2) the coefficient of viscosity in cosmology is given by: $\zeta =  \frac{\rho \left< c \right>}{3f k_\mathrm{B}T_\mathrm{m}} m\lambda_\mathrm{m}$ which may depend on the cosmic time (in general). Therefore the coefficient of viscosity may not be a constant, rather it can vary with the cosmic expansion of the universe.

The above arguments clearly indicate the importance of the bulk viscosity of fluid during the cosmological evolution of the universe, as the comoving expansion itself generates the viscosity. For instance, the thermodynamic interpretations of viscous cosmology becomes important from its own right. However before going to the thermodynamics, let us discuss the cosmological field equations in presence of bulk viscosity. In presence of the bulk viscosity, the continuity equation of the fluid takes the form as:
\begin{align}
&\, \frac{\partial \rho}{\partial t} = - \nabla\cdot \left[ \left( \rho + p_\mathrm{eff} \right) \bm{v} \right] \nonumber \\
\Rightarrow &\, \frac{\partial \rho}{\partial t} + \left( \rho + p_\mathrm{eff} \right) \left( \nabla\cdot \bm{v} \right) = 0\, , \label{6}
\end{align}
with $\rho$ and $p_\mathrm{eff}$ are the energy density and the total pressure of the viscous fluid respectively (here $p_\mathrm{eff}$ contains the viscous pressure, see Eq.~(\ref{Rev-3})). Owing to $\bm{v} = dH \bm{r}$ (in spherical coordinate), one gets $\nabla\cdot \bm{v} = \frac{1}{d^2} \frac{\partial}{\partial d} \left[ d^2 \left( dH\right) \right] = 3H$. As a result, Eq.~(\ref{6}) leads to the following conservation equation for the fluid inside the horizon,
\begin{align}
\dot\rho + 3 H \left( \rho + p_\mathrm{eff} \right) = 0\, .
\label{conservation eqn}
\end{align}
Consequently, the effective equation of state (EoS) of the viscous fluid is defined by:
\begin{eqnarray}
 \omega_\mathrm{eff} = p_\mathrm{eff}/\rho \, .
 \label{Rev-1}
\end{eqnarray}
With the viscous pressure $p_\mathrm{vis} = -3\zeta H$, the effective EoS can be decomposed by,
\begin{eqnarray}
 \omega_\mathrm{eff} = \omega + \frac{p_\mathrm{vis}}{\rho} = \omega - \frac{3\zeta H}{\rho} \, ,
 \label{Rev-2}
\end{eqnarray}
where $\omega$ is the EoS parameter in absence of viscosity of the fluid, and the second term in the rhs of the above equation is solely due to the viscosity. Therefore in presence of the bulk viscosity, $\omega_\mathrm{eff}$ represents the EoS parameter of the viscous fluid and $\omega$ acts as a parameter which has the same value of the ideal EoS parameter of the fluid (i.e., $\omega = 1/3$ for radiation, or, $\omega = 0$ for dust etc.) \cite{Padmanabhan:1987dg}. Eqs.~(\ref{Rev-1}) and (\ref{Rev-2}) immediately lead to the following expression of $p_\mathrm{eff}$ as,
\begin{eqnarray}
 p_\mathrm{eff} = \omega \rho - 3\zeta H = p - 3\zeta H \, ,
 \label{Rev-3}
\end{eqnarray}
where we define $p = \omega \rho$.  Thus as a whole --- $(\rho, p_\mathrm{eff})$ represent the energy density and the total pressure (or the effective pressure) of the viscous fluid, respectively, with the effective EoS is given by $\omega_\mathrm{eff}$; while $p = \omega \rho$ is that part of the $p_\mathrm{eff}$, which corresponds to the ideal EoS counterpart of the viscous fluid. Therefore $p$, for the viscous fluid, has the same functional form $p = p(\rho)$ as in the case of without viscosity (see the discussions after Eq.(49.6) in \cite{Landau}). Moreover $\omega_\mathrm{eff} \rightarrow \omega$, and thus $p_\mathrm{eff} \rightarrow p$, in the limit of $\zeta \rightarrow 0$.

With Eq.~(\ref{Rev-3}), the energy-momentum tensor of such a viscous fluid has the following expression \cite{PhysRevD.79.103521,BARROW1986335,PhysRevD.101.044010,doi:10.1142/S0218271817300245,PhysRevD.73.043512,PhysRevD.72.023003,Fabris_2006,PhysRevD.82.063507,Avelino_2009,Avelino_2010}:
\begin{eqnarray}
 T_\mathrm{\mu\nu} = \left(\rho + p - 3\zeta H\right)U_\mathrm{\mu}U_\mathrm{\nu} + \left(p - 3\zeta H\right)g_\mathrm{\mu\nu} \ ,
 \label{EM-tensor}
\end{eqnarray}
which indeed maintains the symmetry of the underlying spacetime given by Eq.~(\ref{metric}). Here $U_\mathrm{\mu} = \left\{1,0,0,0\right\}$ is the four velocity of the fluid with respect to a comoving observer. The coefficient of viscosity $\zeta$ is generally considered to be a constant, however as mentioned earlier, it can vary with the cosmic expansion. In the present context, our main concern is to examine the thermodynamic consequences of viscous cosmology. Thus we do not put any constraint on the coefficient of viscosity, and take $\zeta$ to be of a very general form, i.e. $\zeta = \zeta(H,\dot{H},...)$. Actually we will show that our findings regarding the thermodynamic interpretations of viscous cosmology hold for general $\zeta$. With the above energy-momentum tensor, the FLRW equations become,
\begin{eqnarray}
 H^2&=&\frac{8\pi G}{3}\rho \ ,\nonumber\\
\dot{H}&=&-4\pi G\left(\rho + p - 3\zeta H\right) \ ,
 \label{FLRW eqn}
\end{eqnarray}
(with $G$ being the Newton's gravitational constant). Clearly Eq.~(\ref{FLRW eqn}) and Eq.~(\ref{conservation eqn}) are not independent to each other, actually the conservation equation can be obtained from the other two. In the present work, we consider that the fluid satisfies the null energy condition given by: $1 + \omega_\mathrm{eff} > 0$ which, due to Eq.~(\ref{FLRW eqn}), boils down to the following condition:
\begin{eqnarray}
 \frac{8\pi G\zeta}{H} < 1 + \omega \, ,
 \label{condition}
\end{eqnarray}
and moreover, provides $\dot{H} < 0$, i.e. the Hubble parameter decreases with time. The above condition (\ref{condition}) also needs to be satisfied in order to have a ghost free scalar perturbation variable over a spatially flat FLRW spacetime \cite{DeFelice:2009bx}.

\section{Thermodynamics of apparent horizon in viscous cosmology}
It is well known that the Bekenstein-Hawking like entropy for the apparent horizon leads to the usual FLRW equations (i.e. without viscosity) from the first thermodynamic law of the horizon. However due to the presence of the viscosity in the present context, the horizon entropy (that leads to the FLRW Eq.~(\ref{FLRW eqn})) gets modified by the coefficient $\zeta$. In the present section, we will determine the proper form of the entropy of the apparent horizon in presence of bulk viscosity.

Differentiating the first one of Eq.~(\ref{FLRW eqn}),
\begin{eqnarray}
 -\left(\frac{2}{R_\mathrm{h}^3}\right)dR_\mathrm{h} = \frac{8\pi G}{3}d\rho \ ,
 \label{en-1}
\end{eqnarray}
which, owing to the conservation law of the fluid, takes the following form,
\begin{eqnarray}
 T_\mathrm{h}\left(\frac{2\pi R_\mathrm{h}}{G}\right)dR_\mathrm{h} = 4\pi R_\mathrm{h}^2\left(\rho + p - 3\zeta H\right)\left(1 - \frac{\dot{R}_\mathrm{h}}{2}\right) \ .
 \label{en-2}
\end{eqnarray}
The internal energy of the fluid inside the horizon is given by: $E = \rho V$, with $V = \frac{4\pi}{3}R_\mathrm{h}^3$ representing the volume enclosed by the apparent horizon. As a consequence, one may derive the following identity:
\begin{eqnarray}
 4\pi R_\mathrm{h}^2\left(\rho + p - 3\zeta H\right)dt = -dE + \rho dV \ ,
 \label{en-3}
\end{eqnarray}
where, once again, the conservation of the fluid has been used. Due to the above expression, Eq.~(\ref{en-2}) turns out to be,
\begin{eqnarray}
 T_\mathrm{h}\left(\frac{2\pi R_\mathrm{h}}{G}\right)dR_\mathrm{h} = \left(-dE + \rho dV\right)\left(1 - \frac{\dot{R}_\mathrm{h}}{2}\right) \ .
 \label{en-4}
\end{eqnarray}
By using the FLRW Eq.~(\ref{FLRW eqn}) and $dV = -\frac{4\pi}{H^4}dH$ (the differential change of the volume enclosed by the horizon), Eq.~(\ref{en-4}) is written as,
\begin{eqnarray}
 T_\mathrm{h}\left\{\frac{2\pi R_\mathrm{h}}{G} - \frac{12\pi^2\zeta}{H^2\left(1 + \frac{\dot{H}}{2H^2}\right)}\right\}dR_\mathrm{h} = -dE + \frac{1}{2}\left(\rho - p\right)dV \ ,
 \label{en-5}
\end{eqnarray}
Now, if we may identify the entropy of the apparent horizon to be of the form like,
\begin{eqnarray}
 S_\mathrm{h} = \left(\frac{A}{4G}\right) - 12\pi^2\int \frac{\zeta}{H^2\left(1 + \frac{\dot{H}}{2H^2}\right)} dR_\mathrm{h} \ ,
 \label{hor-entropy}
\end{eqnarray}
then Eq.~(\ref{en-5}) designates the thermodynamic law of the apparent horizon, and is given by:
\begin{eqnarray}
 T_\mathrm{h}dS_\mathrm{h} = -dE + \frac{1}{2}\left(\rho - p\right)dV \ .
 \label{hor-law}
\end{eqnarray}
Thus as a whole, the horizon entropy and the corresponding thermodynamic law in the context of viscous cosmology are given by Eq.~(\ref{hor-entropy}) and Eq.~(\ref{hor-law}) respectively, where $\zeta = \zeta(H,\dot{H},...)$ is considered to have a general form. Here it may be noted that for $\zeta = 0$, i.e. without the viscosity, the horizon entropy is reduced to the Bekenstein-Hawking entropy, as expected. The thermodynamic law of Eq.~(\ref{hor-law}), or equivalently the form of $S_\mathrm{h}$ in Eq.~(\ref{hor-entropy}), can be simplified to obtain (by using the FLRW Eq.~(\ref{FLRW eqn})),
\begin{eqnarray}
 T_\mathrm{h}\frac{dS_\mathrm{h}}{dt} = \frac{3}{8G}\left(1 - 3\omega\right)\left\{1 + \omega - \frac{8\pi G \zeta}{H}\right\} \ .
 \label{heat-hor}
\end{eqnarray}
The above equation clearly indicates that, due to the same condition shown in Eq.~(\ref{condition}), the horizon entropy monotonically increases with time, in particular,
\begin{eqnarray}
 dS_\mathrm{h} > 0 \, ,
 \label{increase-hor-entropy}
\end{eqnarray}
provided $-1 < \omega < 1/3$, i.e. the fluid satisfies the null energy condition and tends to radiation as $\omega \rightarrow 1/3$.

\section{Thermodynamics of the fluid inside the horizon and the second law of thermodynamics}

The fluid inside the horizon exhibits a viscous like pressure, generated due to the comoving expansion of the universe, and having the energy momentum tensor given by Eq.~(\ref{EM-tensor}) where $\zeta = \zeta(H,\dot{H},...)$ is the coefficient of viscosity. Due to the difference between the comoving expansion speed of the universe ($v_\mathrm{c}$) and the speed of formation of the apparent horizon ($v_\mathrm{h}$), the fluid exhibits a flux through the horizon and thus the fluid behaves as an open system \cite{Odintsov:2024ipb}. In particular, the comoving expansion speed at a physical distance $D$ from a comoving observer is $v_\mathrm{c} = HD$, while $v_\mathrm{h} = -\dot{H}/H^2$. Therefore, the fluid inside the horizon obeys the thermodynamics similar to an open system, i.e.,
\begin{eqnarray}
 T_\mathrm{m}dS_\mathrm{m} = (\mathrm{increase~of~internal~energy}) + (\mathrm{work~done}) + (\mathrm{energy~flux~through~horizon}) \, ,
 \label{thermo-m-1}
\end{eqnarray}
where $T_\mathrm{m}$ and $S_\mathrm{m}$ represent the temperature and the entropy for the fluid, respectively. It may be noted that the fluid's temperature, in general, is considered to be different than the horizon temperature --- later we will show that the viscous cosmology demands $T_\mathrm{m} \neq T_\mathrm{h}$. Now let us individually determine all the terms present in the rhs of Eq.~(\ref{thermo-m-1}).
\begin{itemize}
 \item \underline{Internal energy}: The total internal energy of the fluid inside the horizon, i.e. enclosed by the volume $V = \frac{4\pi}{3}R_\mathrm{h}^3$, is given by $E = \rho V$ (at an instant $t$); therefore the change of internal energy during $dt$ comes as:
 \begin{eqnarray}
 dE = \rho dV - 3H\left(\rho + p - 3\zeta H\right)Vdt \, ,
 \label{thermo-m-2}
\end{eqnarray}
where we use the conservation law for the matter field to arrive at the above expression.

\item \underline{Work done}: The work density by the matter fields is defined by the projection of the corresponding energy-momentum tensor along the normal of the apparent horizon. In particular, $dW = \frac{1}{2}T_\mathrm{ab}h^{ab}dV$ (where $h^{ab}$ is the induced metric along the normal of the horizon) which, due to the spatially flat FLRW metric, takes the following form:
\begin{eqnarray}
 dW = \frac{1}{2}\left(p - 3\zeta H - \rho\right)dV \, .
 \label{thermo-m-3}
\end{eqnarray}

\item \underline{Flux term}: In this regard, we need to realize that the comoving expansion speed of the universe ($v_\mathrm{c}$) differs from the speed of the formation of the apparent horizon ($v_\mathrm{h}$), which in turn results to a flux of the matter fields through the horizon. At an instant $t$, the radius of the apparent horizon is $R_\mathrm{h}(t) = \frac{1}{H(t)}$, then after a time $dt$, the horizon goes to the radius of $R_\mathrm{h}(t+dt) = \frac{1}{H(t+dt)} = \frac{1}{H(t)} - \frac{\dot{H}}{H^2}dt$ (in the leading order of $dt$). On other hand, the comoving surface, that coincides with the horizon at time $t$, moves to the radius $R_\mathrm{c}(t+dt) = \frac{1}{H(t)} + dt$ at the time $t+dt$ (recall that $v_\mathrm{c}(t) = 1$, i.e. the comoving surface coinciding with the horizon at time $t$ has unit speed). Therefore,
\begin{eqnarray}
 V_\mathrm{c}(t+dt) - V(t+dt) = \frac{4\pi}{3}\left(R_\mathrm{c}^3 - R_\mathrm{h}^3\right) = \frac{4\pi}{H^2}\left(1 + \frac{\dot{H}}{H^2}\right)dt \, ,
 \label{thermo-m-4}
\end{eqnarray}
depicting the gap between the comoving volume and the volume enclosed by the apparent horizon (i.e. the visible universe) at time $t+dt$. It may be noted that during an accelerating stage of the universe, when $-\frac{\dot{H}}{H^2} < 1$, the comoving expansion speed is larger than the speed of apparent horizon and thus $V_\mathrm{c}(t+dt) > V(t+dt)$; while for a decelerating stage, the reverse scenario occurs. Eq.~(\ref{thermo-m-4}) immediately determines the outward flux of the fluid through the horizon as,
\begin{eqnarray}
 d\mathcal{F} = \left(\rho + 3p_\mathrm{eff}\right)\times\left[V_\mathrm{c}(t+dt) - V(t+dt)\right] = -\frac{2\pi\rho}{H^2}\left(1 + 3\omega - \frac{24\pi G \zeta}{H}\right)^2dt \, .
 \label{thermo-m-5}
\end{eqnarray}
In spirit of the matter flux, the effective energy density $(\rho + 3p_\mathrm{eff})$ can be realized as the chemical potential ($\mu$) of the viscous fluid, in particular, $\mu \propto -(\rho + 3p_\mathrm{eff})$ which depicts that the chemical potential is negative for $(\rho+3p_\mathrm{eff}) > 0$ while $\mu > 0$ for $(\rho+3p_\mathrm{eff}) < 0$. This is expected as the universe with $(\rho+3p_\mathrm{eff}) > 0$ undergoes through a deceleration era during when the fluid experiences an attractive gravitational force, that in turn results to $\mu < 0$; while the fluid with $(\rho+3p_\mathrm{eff}) < 0$ results to an accelerating universe and experiences a kind of repulsive force that shows up through a positive valued chemical potential.
\end{itemize}

By using the expressions from Eqs.~(\ref{thermo-m-2}), (\ref{thermo-m-3}) and (\ref{thermo-m-5}), we obtain the thermodynamic law for the viscous fluid inside the horizon from Eq.~(\ref{thermo-m-1}) as follows:
{\small{\begin{eqnarray}
 T_\mathrm{m}\frac{dS_\mathrm{m}}{dt}=-\frac{3}{8G}\left\{\left(1 - 3\omega\right)\left(1 + \omega - \frac{8\pi G \zeta}{H}\right) + 2\left(1 + 3\omega - \frac{24\pi G \zeta}{H}\right)^2\right\} - \frac{9\pi\zeta}{H}\left(1 + \omega - \frac{8\pi G \zeta}{H}\right) \, .
 \label{thermo-m-7}
\end{eqnarray}}}
This, due to Eq,~(\ref{condition}), depicts that $T_\mathrm{m}\dot{S}_\mathrm{m} < 0$, i.e. the entropy of the fluid monotonically decreases with the cosmic time.

Therefore it is clear from Eq.~(\ref{increase-hor-entropy}) and Eq.~(\ref{thermo-m-7}) that the entropy of the horizon increases while the fluid's entropy decreases with the expansion of the universe, in particular,
\begin{eqnarray}
 T_\mathrm{h}\frac{dS_\mathrm{h}}{dt} > 0~~~~~\mathrm{and}~~~~~T_\mathrm{m}\frac{dS_\mathrm{m}}{dt} < 0 \, .
 \label{exc-1}
\end{eqnarray}
This in turn indicates that the heat energy is released by the fluid and gets absorbed by the apparent horizon --- thus the flow of heat energy is directed from the fluid towards the horizon. Such spontaneous flow of heat energy demands,
\begin{eqnarray}
 T_\mathrm{m} \geq T_\mathrm{h} \, ,
 \label{exc-2}
\end{eqnarray}
i.e. there are two possibilities in the present context: either $T_\mathrm{m} = T_\mathrm{h}$ or $T_\mathrm{m} > T_\mathrm{h}$. In the next two subsections we will examine which, out of these two possibilities, is allowed during the entire cosmic evolution of the universe started from inflation to the dark energy era.

Since the apparent horizon absorbs heat energy during the cosmic time, let us consider that the horizon entropy increases by an amount $\Delta S_\mathrm{h}$ within time $\Delta t$. During the increment of $\Delta S_\mathrm{h}$, the change of the Hubble parameter can be obtained by using Eq.~(\ref{hor-entropy}), and is given by:
\begin{eqnarray}
 dH = -\left(\frac{GH^3}{2\pi}\right)\left\{1 - \frac{6\pi G \zeta}{H\left(1 + \frac{\dot{H}}{2H^2}\right)}\right\}^{-1} \Delta S_\mathrm{h} \ .
 \label{change-Hubble}
\end{eqnarray}

\subsection*{Reversible ($T_\mathrm{m} = T_\mathrm{h}$) or Irreversible ($T_\mathrm{m} > T_\mathrm{h}$) cases of the second law of thermodynamics}

In this subsection, we will examine whether the expansion of the universe allows the possibility of $T_\mathrm{m} = T_\mathrm{h}$. In this case, the apparent horizon is in thermal equilibrium with the fluid and thus the heat exchange from the fluid to the horizon is reversible in nature. Therefore the change of total entropy vanishes, i.e.,
\begin{eqnarray}
 \Delta S_\mathrm{h} + \Delta S_\mathrm{m} = 0 \, .
 \label{sub-1-1}
\end{eqnarray}
Let $\left|\Delta Q_\mathrm{m}\right|$ is the amount of heat released by the fluid during the time, by which, the horizon entropy increases by $\Delta S_\mathrm{h}$, then
\begin{eqnarray}
 \left|\Delta Q_\mathrm{m}\right| = T_\mathrm{m}\left|\Delta S_\mathrm{m}\right| \, ,
 \label{sub-1-2}
\end{eqnarray}
which, due to $T_\mathrm{m} = T_\mathrm{h}$ along with Eq.~(\ref{sub-1-1}), can be written as:
\begin{eqnarray}
 \left|\Delta Q_\mathrm{m}\right| = T_\mathrm{h}\Delta S_\mathrm{h} \, .
 \label{sub-1-3}
\end{eqnarray}
Eq.~(\ref{sub-1-3}) designates the reversible case of the second law of thermodynamics. Owing to Eq.~(\ref{thermo-m-7}), the above expression leads to,
{\small{\begin{eqnarray}
 \left[\frac{3}{8G}\left\{\left(1 - 3\omega\right)\left(1 + \omega - \frac{8\pi G \zeta}{H}\right) + 2\left(1 + 3\omega - \frac{24\pi G \zeta}{H}\right)^2\right\} + \frac{9\pi\zeta}{H}\left(1 + \omega - \frac{8\pi G \zeta}{H}\right)\right]\Delta t = T_\mathrm{h}\Delta S_\mathrm{h} \, ,
 \label{sub-1-4}
\end{eqnarray}}}
where $\Delta t$ is the time interval during when the horizon entropy increases by $\Delta S_\mathrm{h}$. As a consequence, Eq.~(\ref{change-Hubble}) immediately leads to $\Delta t$ as,
\begin{eqnarray}
 \Delta t = \frac{dH}{\dot{H}} = -\left(\frac{GH^3}{2\pi\dot{H}}\right)\left\{1 - \frac{6\pi G \zeta}{H\left(1 + \frac{\dot{H}}{2H^2}\right)}\right\}^{-1} \Delta S_\mathrm{h} \, .
 \label{sub-1-5}
\end{eqnarray}
By using the above expression of $\Delta t$, along with the FLRW Eq.~(\ref{FLRW eqn}) and $T_\mathrm{h} = \frac{H}{2\pi}\left(1 + \frac{\dot{H}}{2H^2}\right)$, Eq.~(\ref{sub-1-4}) finally boils down to the following condition:
\begin{eqnarray}
 \left(1 + 3\omega - \frac{24\pi G \zeta}{H}\right)^2 + \frac{12\pi G \zeta}{H}\left(1 + \omega - \frac{8\pi G \zeta}{H}\right) = 0 \, .
 \label{sub-1-6}
\end{eqnarray}
Therefore the above condition needs to be hold in order to satisfy the reversible case of the second law of thermodynamics. However, as both the terms in the lhs of Eq.~(\ref{sub-1-6}) are positive (recall Eq.~(\ref{condition})), the above condition is not satisfied at any instant of the cosmic expansion. In order to realize this more, it may be noted that the first term in the lhs of Eq.~(\ref{sub-1-6}) is related to the acceleration term of the universe, in particular, the above equation can be expressed in the following manner:
\begin{eqnarray}
 \left(\frac{\ddot{a}}{aH^2}\right)^2 + \frac{3\pi G \zeta}{H}\left(1 + \omega - \frac{8\pi G \zeta}{H}\right) = 0 \, .
 \label{sub-1-7}
\end{eqnarray}
This clearly demonstrates that the cosmic expansion of the universe, in presence of bulk viscosity of the fluid, does not allow the reversible case of the second law of thermodynamics at any instant of $\ddot{a} > 0$ or $\ddot{a} < 0$ or $\ddot{a} = 0$. This is unlike to the scenario of without viscosity where the reversibility holds for $\ddot{a} = 0$ \cite{Odintsov:2024ipb}, i.e. when the universe expands with no acceleration or no deceleration (for instance, the transitions from early inflation to standard Big-Bang cosmology (SBBC) and from SBBC to the late dark energy era). Actually in the context of viscous cosmology, the appearance of $\zeta$ in the second term of lhs of Eq.~(\ref{sub-1-7}) is the sole reason to not hold the above condition at $\ddot{a} = 0$. Thereby we may argue that due to the presence of bulk viscosity, the expansion of the universe loses the thermodynamic reversibility compared to that of in absence of viscosity.

We now focus to examine the second possibility $T_\mathrm{m} > T_\mathrm{h}$, in which case, the fluid and the apparent horizon are not in thermal equilibrium, and the change of total entropy is positive, i.e.,
\begin{eqnarray}
 \Delta S_\mathrm{h} + \Delta S_\mathrm{m} > 0 \, .
 \label{sub-2-1}
\end{eqnarray}
Once again, if $\left|\Delta Q_\mathrm{m}\right|$ is the amount of heat released by the matter fields (within the interval when the horizon entropy increases by $\Delta S_\mathrm{h}$), then we have the following inequality,
\begin{eqnarray}
 \left|\Delta Q_\mathrm{m}\right| > T_\mathrm{h}\Delta S_\mathrm{h}\left(\frac{\left|\Delta S_\mathrm{m}\right|}{\Delta S_\mathrm{h}}\right) \, ,
\label{sub-2-3}
\end{eqnarray}
which, due to Eq.~(\ref{sub-2-1}), gets automatically satisfied if,
\begin{eqnarray}
 \left|\Delta Q_\mathrm{m}\right| > T_\mathrm{h}\Delta S_\mathrm{h} \, .
 \label{sub-2-4}
\end{eqnarray}
This is the irreversible expression of the second law of thermodynamics. By following the same procedure (as in the case of reversibility), the above expression turns out to be,
\begin{eqnarray}
 \left(1 + 3\omega - \frac{24\pi G \zeta}{H}\right)^2 + \frac{12\pi G \zeta}{H}\left(1 + \omega - \frac{8\pi G \zeta}{H}\right) > 0 \, ,
 \label{sub-2-6}
\end{eqnarray}
which, in terms of the acceleration term, can be written as,
\begin{eqnarray}
 \left(\frac{\ddot{a}}{aH^2}\right)^2 + \frac{3\pi G \zeta}{H}\left(1 + \omega - \frac{8\pi G \zeta}{H}\right) > 0 \, .
 \label{sub-2-7}
\end{eqnarray}
This clearly demonstrates that the presence of $\zeta$ makes the above inequality true for $\ddot{a} > 0$ as well as for $\ddot{a} < 0$ or $\ddot{a}=0$. In this regard, the coefficient of viscosity is significantly important for $\ddot{a} = 0$, in which case, the above condition does not hold true without the presence of $\zeta$.  Therefore the irreversible case of the second law of thermodynamics, in the context of viscous cosmology, gets satisfied during the entire cosmic evolution of the universe. This is unlike to the scenario of without viscosity where the irreversibility fails for $\ddot{a} = 0$ \cite{Odintsov:2024ipb}. Thereby comparing the two scenarios, i.e. with / without the bulk viscosity of the fluid (although the viscosity, in general, is inevitable in cosmological context, as we show that the comoving expansion of the universe itself generates the viscosity, see Sec.~[\ref{sec-generation}]), we may argue that the presence of viscosity makes the expansion of the universe more irreversible than that of without the viscosity.

\section{Conclusion}
In the present work, we first propose a novel mechanism for the generation of bulk viscosity in cosmological context, and then address the thermodynamic implications of viscous cosmology. In particular, we show that the velocity gradient of the comoving expansion of the universe (along the distance measured from a comoving observer) in turn leads to a viscous like pressure of the fluid inside the apparent horizon. This relies on the consideration that the fluid obeys the principles of the kinetic theory of gas, having a temperature $T_\mathrm{m}$ and a mean free path $\lambda_\mathrm{m}$. The coefficient of viscosity is found in terms of the mean free path and the temperature of the fluid, which shows that the coefficient of viscosity in cosmology may depend on the cosmic time (in general). Such mechanism of the bulk viscosity immediately indicates the importance of the viscosity of the fluid during the cosmological evolution of the universe, as the comoving expansion itself generates the viscosity. Therefore the thermodynamic implications of viscous cosmology becomes important from its own right. In this regard, we show that the cosmological evolution of the universe with a bulk viscosity $naturally$ satisfies the first and the second laws of thermodynamics of the apparent horizon, without imposing any exotic condition and for a general form of the coefficient of viscosity. By using the first thermodynamic law of the apparent horizon, we determine the entropy of the horizon corresponding to the field equations of the viscous cosmology. Such horizon entropy deviates from the usual Bekenstein-Hawking entropy, and the deviation is proportional to the coefficient of viscosity ($\zeta$). Importantly, it turns out that the entire cosmic evolution of the universe, in presence of the bulk viscosity of the fluid, naturally admits the $irreversible$ case of the second law of thermodynamics; and this argument holds true irrespective of any form of $\zeta$. This affirms the thermodynamic correspondence of viscous cosmology. Regarding the second law of thermodynamics, the present scenario is different compared to the scenario of without viscosity where the cosmic expansion of the universe allows both the reversible and irreversible cases of the second law of thermodynamics. Thereby comparing the two scenarios, i.e. with / without the bulk viscosity of the fluid, we may argue that the presence of viscosity makes the expansion of the universe more irreversible than that of without the viscosity.

\bibliography{Landauerpr-vis-cosmology}
\bibliographystyle{./utphys1}

\end{document}